\documentclass[graybox]{svmult}
\usepackage{amsmath}
\usepackage{amsfonts}
\usepackage{amssymb}
\usepackage{graphicx}
\usepackage{enumerate}
\usepackage{algorithm}
\usepackage{comment}
\usepackage{multirow}
\usepackage{algorithmic, amsmath}
\usepackage{calrsfs}
\DeclareMathAlphabet{\pazocal}{OMS}{zplm}{m}{n}
\usepackage{longtable}

\usepackage{cite} 
\usepackage{times}
\usepackage{subfig}
\usepackage{url}
\usepackage{cite}
\usepackage{multicol}        % used for the two-column index
\usepackage[bottom]{footmisc}
\makeindex  
\begin{document}
\begin{frontmatter}

\title*{Impact of Distance on Epidemiological Dynamics in Human Connection Network with Mobility}
\author{Md. Arquam $^1$, Suchi Kumari$^2$, Utkarsh Tiwari$^2$, Mohammad Al-saffar$^3$}
%Mohamed A.G. Hazber

\authorrunning{Md. Arquam}
%% (feature abused for this document to repeat the title also on left hand pages) 
%
%% the affiliations are given next; don't give your e-mail address
%% unless you accept that it will be published
\institute{$^1$ Department of Computer Science and Engineering, Indian Institute of Information Technology, Sonepat, Haryana, India, Email: md.arquam@iiitsonepat.ac.in. \\ 
$^2$ Department of Computer Science and Engineering, Shiv Nadar Institute of Eminence, Delhi-NCR, Uttar Pradesh, India, Email: suchi.singh24@gmail.com, ut353@snu.edu.in. \\
$^3$ School of Computer Science and Engineering, Hail University, KSA, Email: m.alsaffar@uoh.edu.sa.}

\maketitle
\vspace*{-.25cm}
\abstract{: The spread of infectious diseases is often influenced by human mobility across different geographical regions. Although numerous studies have investigated how diseases like SARS and COVID-19 spread from China to various global locations, there remains a gap in understanding how the movement of individuals contributes to disease transmission on a more personal or human-to-human level. Typically, researchers have employed the concept of metapopulation movement to analyze how diseases move from one location to another. This paper shifts focus to the dynamics of disease transmission, incorporating the critical factor of distance between an infected person and a healthy individual during human movement. The study delves into the impact of distance on various parameters of epidemiological dynamics throughout human mobility. Mathematical expressions for important epidemiological metrics, such as the basic reproduction number ($R_0$) and the critical infection rate ($\beta_{critical}$), are derived in relation to the distance between individuals. The results indicate that the proposed model closely aligns with observed patterns of COVID-19 spread based on the analysis done on the available datasets.}

\end{frontmatter}

\section{Introduction}\label{sec:Introduction} 
Complex networks are powerful tools for modeling real-world systems, with applications in areas like biological network \cite{chen2024simplicial}, social networks \cite{kumar2021quantum,kumari2024novel}, communication networks \cite{kumari2017modeling}, and so on. They are particularly useful for studying the spread of information or diseases within a network. The major tasks include identifying nodes that can maximize information spread, minimize rumors, or reveal superspreaders in epidemics or computer viruses. Previous approaches require understanding the spreading abilities of nodes, often involving costly simulations like Monte Carlo methods in propagation models such as SIR. 
These models aim to understand epidemic dynamics and provide accurate predictions of spread over time and across locations, whether nationally or globally \cite{cooper2020sir}. Most models fall under the category of ``compartmental models" where the population is divided into compartments representing healthy individuals, the infected, the ill, the deceased, the recovered, and others \cite{purkayastha2021comparison}. The transitions between these compartments are governed by a system of interconnected ordinary differential equations (ODEs) \cite{reyne2022principles}.

Many epidemic models operate on the fundamental assumption of population homogeneity, implying that everyone has an equal chance of encountering someone new.  This assumption is valid for local epidemics confined to a single community, where it allows the model to be simplified into mean field equations. However, this approach no longer applies to larger populations. Individuals who commute between locations play a crucial role in spreading diseases on a national and global scale, acting as infection carriers in new areas \cite{balcan2009multiscale}. This process, which is influenced by movement and interaction across regions, cannot be adequately captured by traditional compartmental models. In their research, the authors explored the behavior of positive solutions to a reaction-diffusion system with Neumann boundary conditions, which is a more suitable approach for modeling such phenomena.

Recent advancements include network models that deviate from the assumption of homogeneous mixing \cite{silva2023epidemic}. These models use ordinary differential equations (ODEs) by averaging encounters into grouped contagion rate coefficients, accounting for well-defined groups with either homogeneous or heterogeneous mixing conditions \cite{arquam2020impact,arquam2020integrating}. Traditionally, analytical studies on epidemic dissemination have predominantly focused on compartmental epidemic models grounded in ordinary differential equations. In scale-free networks, Liu \textit{et al.} \cite{liu2011epidemic} presented an SEIRS epidemic model in which the number of active contacts at each vertex was either constant or proportional to its degree. It is important to note that contemporary research often employs various differential equations to study the complex dynamics of epidemic spread. These are the most popular ways of studying and examining the complex dynamic system of epidemic spread. When looking at how epidemic spread, the dynamics of their distribution must be studied. In the event of an epidemic spreading from one location to another, geometric parameters must be examined.

This paper focused on the geometric distance and its impact on epidemic spreading in human connections and networks with mobility. Hence, this paper's objectives are:
\begin{itemize}
\item Development of Novel Approach: Develops a method to differentially parameterized SIR models of epidemics to geometric distance parameters.

\item Distance Analysis of Epidemic Spread: Study the geometric distances in people and how it affects the spread of diseases.

\item Framework Model Development: Create formulas for epidemiological data like the basic reproduction number $R_0$ and critical infection rate $\beta_{critical}$ based on distance.

\item Human Mobility Modeling in Epidemic Spread: Integrate human movement and migration flows concerning disease spread in different places.

\item Real Data Model Assessment: Test the model with real data from the COVID-19 pandemic and compare it with infection dynamics observed in real life.

\end{itemize}

The paper is organised as follows: Section 2 offers a comprehensive review of the existing literature and relevant research on epidemic spreading. Section 3 presents a detailed description of the proposed model, including its key components and processes. The dynamics of epidemic transmission are explored, and the basic reproduction number is calculated to aid in understanding how epidemics propagate within the network. Section 4 discusses the simulation results and analyzes the outcomes, providing insights into potential scenarios and their impacts. Finally, Section 5 summarizes the findings and highlights possible directions for future research and development in this area. This structure enables a thorough examination of the epidemic spreading model within the SIR framework, focusing on geometric distance and its influence.

%SIR and SIS epidemic models with bilinear incidence and migration between two patches were taken into consideration by Yang et al. \cite{yang2010global}, wherein diseased people are prevented from moving from one patch to another by medical screening.The thresholds for categorising the global dynamics of the models in terms of the model parameters were discovered, and they were able to determine the global asymptotic stability of the equilibrium between the presence and absence of disease.The fundamental drawback of compartmental models, according to Wolkewitz and Schumacher \cite{wolkewitz2011simulating}, is that a number of their parameters are reliant on educated assumptions from experts (default values) rather than being determined from the study's data. Lee et al. \cite{lee2012effect} expanded the SEIR model to take into account population movement across cities and looked into how well travel restrictions worked as a preventative measure. 

% To integrate age structures, for example, first-order partial differential equations are used; second-order partial differential equations are appropriate when a diffusion term is present; and integral differential equations or differential equations are frequently used when time delay or delay factors are taken into account.

\section{Literature survey}

This section reviews related research in the field of epidemic spreading for various diseases. Samsuzzoha \textit{et al.} \cite{samsuzzoha2012numerical} developed a diffusive epidemic model to describe the spread of influenza. They introduced a vaccinated diffusive compartmental model to analyze the impact of vaccination and diffusion on influenza transmission dynamics. Epidemics often spread through the movement of infected individuals between locations. To model this, some metapopulation models simulate the effects of migration, linking local communities through a network of places or ``patches" and using a transfer matrix to describe interactions between them \cite{brauer2017mathematical}. Most of these models focus on the global spread of diseases, assuming unidirectional travel and capturing worldwide transmission dynamics \cite{kirkeby2021practical,baker2022infectious}.

Within the category of metapopulation models, both stochastic and deterministic types are commonly used, differing in their level of resolution. Some models, for example, focus only on the population density at each location. In these cases, population transfer between patches is handled by decreasing the population of one patch and increasing the population of another. Researchers introduce an efficient algorithm for modeling commuting in metapopulation models, which avoids the need for complex individual-level formulations or the expansion of epidemic models to account for a large number of patches \cite{pardo2023epidemic,wang2021final}. Other models, which track the position of everyone, offer greater precision. These models can incorporate important factors that simpler models may overlook, such as typical travel behaviors or the presence of ``super spreaders", individuals who visit multiple locations simultaneously \cite{costa2020outbreak}. However, the main limitation of these detailed models is their high computational cost. 

On a national level, travel patterns differ from random unidirectional movement in terms of both their properties and the effects they have on spreading. Individuals typically commute regularly to the same destination, such as an office or other nearby locations. Reverse commuting also occurs, where people spend part of their time in their ``residence city" but the majority of their day at a different location, such as a workplace \cite{wang2023epidemic}. Many models of metapopulation dynamics focus on unidirectional commuting, assuming that, with high commuting frequency relative to the infection rate, the system will eventually reach a steady state, regardless of the original geographic distribution. However, many models overlook reverse commuting due to the difficulty in tracking the health status of commuters at the end of their workday. When reverse commuting is considered, computational approaches often simulate each commuting group separately \cite{han2023effects}.

Some methods of transmission may exhibit recurring patterns similar to those observed in infectious illnesses, however these patterns may vary in their predictability \cite{bu2022likelihood}. Identifying the key factors contributing to these patterns can provide a framework for assessing treatment efficacy and guiding disease control planning \cite{kumar2022age}. Ghosh \textit{et al.} \cite{ghosh2021optimal} proposed a deterministic compartmental model of infectious disease that incorporates test kits as a key tool for epidemic mitigation. The outbreak size and infection peaks depend highly on targeting nodes with higher degrees or centralities. Csimcsek  \cite{csimcsek2022lexical} introduced a Lexical Sorting Centrality (LSC) for epidemic modeling in complex networks, which is a combination of multiple centrality measures to assess the spreading capabilities of nodes. Using a sorting mechanism like lexical sorting, LSC ranks nodes more efficiently than traditional centrality measures and can better distinguish node spreading abilities with greater accuracy and speed. Contreras \textit{et al.} \cite{contreras2024infection} explored how different contagion models, including simple and complex contagions, influence infection patterns on a network. They found that simple contagions, spreading through individual connections, are robust, while complex contagions are sensitive to model parameters due to their reliance on the interplay between pairwise and group contagions. 

Several studies focus on the analysis of the spread and mitigation of COVID-19. Zino \textit{et al.} \cite{zino2021analysis} reviewed the history and current state of epidemic modeling, analysis, and control in guiding public health decisions during the COVID-19 pandemic. It covers modeling approaches from early scalar differential equations to modern dynamic network models that account for spatial spread and changing human interactions. Goel \textit{et al.} \cite{goel2021mobility} proposed a mobility-based SIR model, considering both global connectivity and local population distribution to better understand the spreading dynamics of pandemics like COVID-19. They considered mixed population as well as real-life interaction in a complex pattern.  Zhu \textit{et al.} \cite{zhu2021connectedness} constructed a complex network of 122 countries to analyze the global COVID-19 infection network considering some parameters such as small-world, close-world, and community structure. The analysis showed that factors such as population density, economic size, exports, and government expenditure increased network connectedness. Countries with higher economic activity and population density were more vulnerable with respect to those with better medical infrastructure had lower connectedness. Johnson  \cite{johnson2024epidemic} explained the importance of network topology on the spreading dynamics of diseases. He explained that the compartmental model generally overlooks the network structure and performs random mixing. But,  this assumption leads to overestimations of both the herd-immunity threshold and basic reproduction number $R_0$. The study suggests that the scale-free networks may be a better model for epidemics like COVID-19 than random mixing. Chen \textit{et al.} \cite{chen2024simplicial} investigated the impact of higher-order social interactions on COVID-19 spread and mitigation. They found that considering group interactions enhances contagion tracking and mitigation strategies, as group members spend more time together.

Li \textit{et al.} \cite{li2022competing} presented a model of two competing simplicial SIS epidemics on a higher-order system represented by a simplicial complex, where infections spread through both pairwise and higher-order (e.g., 2-simplex) interactions. The dynamics of infection depends on tie-strength where the system behaves like a simple graph for a weak infection strength.  As the infection strength increases, an ``alternative dominant" phase emerges, where the survival of one epidemic depends on initial conditions. Fan \textit{et al.} \cite{fan2022epidemics} explored the impact of higher-order interactions in a multilayer network model, where one layer represents a social network and the other a physical contact network. Using a microscopic Markov chain approach and Monte Carlo simulations, the research shows that the 2-simplex interactions in the social network can effectively mitigate epidemic outbreaks, especially when disease transmission rates in the physical network are low or medium.

Silva \textit{et al.} \cite{silva2023epidemic} investigates a SIR model considering local and global epidemic prevalence in heterogeneous networks. It was investigated that local awareness can significantly increase the epidemic threshold, delay peak prevalence, and reduce outbreak size. Li \textit{et al.} \cite{li2021dynamics} proposed a modified signed-SIS model to study epidemic spreading on signed networks, where edges are assigned positive or negative labels that impact transmission rates. They demonstrated that structural balance theory influences epidemic dynamics, affecting peak infection fractions and epidemic thresholds, through analysis on both random and scale-free networks. Rodriguez  \cite{rodriguez2024analysis} analyzed an SIS-type model and examined the influence of network topology on the spreading dynamics of viruses in the network. Through simulations, it shows that regular topologies with low node degrees, like Lattice4, facilitate faster virus extinction compared to denser or less regular topologies. The author eliminated edges and tried to convert the network into a regular graph to induce rapid virus extinction.

\subsection*{Key findings from the literature survey}

Most researchers have considered either the meta-population concept or the radiation model to explain the spreading of infectious diseases from one location to another. But, they failed to explain the connecting pattern of individuals, and their movement from one location to another location. In this course of study, the author has demonstrated the impact of delay in epidemic spreading in human connection network \cite{arquam2019modelling}. Arquam \textit{et al.} \cite{arquam2020epidemic} proposed a model considering change of geometric location and connecting distance of two individuals but failed to explain the impact of connecting distance. Therefore, connecting distance must be analysed while studying the epidemic spreading in human connection network.

\section{Proposed Methodology} \label{PM}

In this section, epidemic spreading due to human movement from one location to other location is explained.
For human movement, a GNMN model \cite{arquam2020epidemic} is taken to consider the human connection network in geometric region.

\begin{table}[]
    \centering
    \caption{Symbols along with their meaning}
    \begin{tabular}{|p{1.5 cm}|p{7.5 cm}|} 
    \hline 
    \textbf{Symbol} & \textbf{Meaning} \\ \hline
     $N$    & Total number of nodes or individuals in the network  \\ \hline
     $E$    & Total number of edges or connections in the network  \\ \hline
     $S_i(t)$ & Number of susceptible individuals at node $i$ at time $t$ \\ \hline  
     $I_i(t)$ & Number of infectious individuals at node $i$ at time $t$ \\ \hline
     $R_i(t)$ & Number of recovered individuals at node $i$ at time $t$ \\ \hline
     $\beta$ & Infection rate \\ \hline
     $\mu$ & Recovery rate \\ \hline
     $d_{ij}$ & Euclidean distance between nodes $i$ and $j$ \\ \hline
     $d_{T}$ & Threshold distance for spreading of infection \\ \hline
     $\sigma$ & Dispersion parameter controlling the effect of distance \\ \hline
     $<k>$ & Average degree of the nodes in the network \\ \hline
     $A$ & Area considered for the node's movement \\ \hline 
     $v_i$ & Speed of node $i$ at time $t$ \\ \hline
    \end{tabular}
    \label{tab:nomenclature}
\end{table}

In the geometric network with mobile nodes (GNMN), the distance between two nodes plays a vital role in the spreading process of diseases in the network. Two nodes cannot infect each other beyond a threshold distance $d_T$. As the spreading of the epidemic depends on the distance, the total number of connections ($E$) can be formulated in Eq. \eqref{e1}.

\begin{equation}
    E = \sum_i \frac{\sum_j d_{ij}}{A} \frac{v_i T}{d_T} \label{e1}
\end{equation}

The first part of the fraction gives the probability of a node $i$ within the vicinity of other nodes in the network. The second component provides the fraction of time a node $i$ is in the active condition in the spreading process.
The average degree ($<k_i>$) of a node $i$ is evaluated considering the distance travelled by the node $i$ within the designated area $A$ and the formulation for a random GNMN is provided in Eq. \eqref{e2}.
\begin{equation}
    <k_i> = N \frac{\sum_j d_{ij}}{A} \label{e2}
\end{equation}

\subsection{SIR model for GNMN}
%\textcolor{red}{Explain some content about SIR model here}
In 1760, Daniel Bernoulli~\cite{bernoulli1760essai} introduced the first mathematical framework for studying the spread of infectious diseases. Building upon this foundation, Kermack and McKendrick~\cite{kermack1927contribution} proposed the classical Susceptible-Infectious-Recovered (SIR) model. The following system of rate equations governs the dynamics of the SIR model:
\vspace{-.4cm}
\begin{eqnarray}
\frac{dS(t)}{dt} & = & -\beta S(t)I(t), \label{RE1} \\
\frac{dI(t)}{dt} & = & \beta S(t)I(t) - \mu I(t), \label{RE2} \\
\frac{dR(t)}{dt} & = & \mu I(t), \label{RE3}
\end{eqnarray}
where $S(t)$, $I(t)$, and $R(t)$ represent the fractions of the susceptible, infected, and recovered populations at time $t$, respectively. The parameters $\beta$ and $\mu$ denote the infection (spreading) rate and recovery rate.

Despite its historical significance, the classical SIR model does not account for heterogeneity or the underlying topology of complex networks, which are critical factors influencing epidemic dynamics in real-world scenarios.
\begin{comment}
    WithWithWithWith
Where $v$ is the velocity of nodes.
Combining the all parameters, the estimated average node degree (⟨k⟩⟨k⟩) in a random mobile geometric network considering distance-based epidemic spreading using a complex network approach in a homogeneous network setting can be expressed as:

\begin{equation} \label{K} 
\langle k \rangle = N\frac{total \enspace distance \enspace traveled \enspace by \enspace a \enspace node}{A}   
\end{equation} \\  
This derivation captures the concept of node mobility, connectivity, and interaction intervals in a random mobile geometric network.
\end{comment}
In this proposed work, SIR Epidemiological Dynamics can be explained using mean-field equation incorporating geometric network topology and change in Susceptible Individuals can be evaluated considering Eq. \eqref{e3}.
    \begin{equation} 
    \frac{dS_i(t)}{dt}  = -\beta \langle k \rangle S_i(t) \sum_{j=1}^N 1_{d_{ij}}\leq d_T e^{\big (-\frac{d_{ij}^2}{2\sigma^2}\big )I_j(t)} \label{e3}
 \end{equation}  
Infection spreading can be defined as,
 \begin{equation}  
   \frac{dI_i(t)}{dt} = \beta \langle k \rangle S_{i}(t)\sum_{j=1}^N1_{d_{ij}\leq d_T}e^{\big (-\frac{d_{ij}^2}{2\sigma^2} \big )I_j(t)} - \mu I_i(t) \label{e4}
  \end{equation}    
    Recovered dynamics can be written as,
     \begin{equation}    \label{R}
    \frac{dR_i(t)}{dt} = \mu I_i(t)  \\
 \end{equation} 
    
The infection model incorporates an indicator function $1_{d_{ij} \leq d_T}$ that ensures only nodes within a threshold distance $d_T$ contribute to the infection process. If the distance $d_{ij}$ between nodes exceeds $d_T$, the function becomes 0, preventing infections from occurring beyond this threshold. The rate of change of infectious individuals at node iii is determined by the product of the infection rate $\beta$, the number of susceptible individuals $S_i(t)$, and the interaction term, which sums over all nodes $j$ within the threshold distance. This interaction is modulated by a Gaussian-like function $e^{-\frac{d_{ij}^2}{2\sigma^2}}$, which reduces the infection probability as the distance between nodes increases. The rate of change of recovered individuals is given by $\mu I_i(t)$, where $\mu$ is the recovery rate and $I_i(t)$ is the number of infectious individuals at node $i$. Finally, the change in susceptible individuals is determined by the decrease in $S_i(t)$ due to infections spreading through the population.

\subsection{Dynamical Behaviour of Epidemiological Model}

The spread of infectious diseases begins when the following conditions must be satisfied,
$I(t) \geq 0$, $S(t) \geq 0$, and $R(t) = 0$. 

\subsubsection{The basic reproduction number ($R_0$)}
The basic reproduction number ($R_0$) represents the average number of secondary infections generated by a single infected individual in a fully susceptible population. It is a key parameter in epidemiology, as it helps assess the potential for disease transmission within a population. The formula for $R_0$ in a Generalized Network Model of Networks (GNMN) incorporates the following key parameters.
\begin{itemize}
    \item Infection rate ($\beta$) and recovery rate ($\mu$): These parameters describe the speed at which the disease spreads and the rate at which individuals recover, respectively.
    \item Network structure ($\langle k \rangle$): The average node degree represents the typical number of connections each node has in the network.
    \item Distance ($d_{ij}$): The inclusion of spatial distance in the epidemic spread means that the distance between individuals in the network influences the likelihood of infection.
\end{itemize}

By considering the aforementioned parameters, the formulation of $R_0$ can be provided in Eq. \eqref{e6}
\begin{equation}
    R_0= \frac{\beta}{\mu}\frac{1}{\langle k \rangle}\sum_{i=1}^Ne^{\big (-\frac{d_{T}^2}{2\sigma^2} \big )}\frac{1}{N}\sum_{j=1}^Ne^{\big (-\frac{d_{ij}^2}{2\sigma^2} \big )} \label{e6}
\end{equation}
 The term $\frac{\beta}{\mu}$ represents the average number of secondary infections caused by a single infected individual before they recover. If $\beta >> \mu$, the network will be rapidly infected, while if $\beta$ is much smaller than $\mu$, the infection will spread more slowly throughout the network. The term $\frac{1}{\langle k \rangle}$ accounts for the average node degree, which reflects the connectivity of the network. The expression $\sum_{i=1}^N e^{(-\frac{d_{T}^2}{2\sigma^2})}$ captures the attenuation effect of the distance threshold, ensuring that nodes beyond a certain distance have a negligible probability of infection. The last expression $\frac{1}{N}\sum_{j=1}^N e^{\left(-\frac{d_{ij}^2}{2\sigma^2}\right)}$ represents the average influence of an infected node $j$ on the infection probability of other nodes, considering the attenuation effect of distance. When $R_0$ exceeds $1$, the disease can spread extensively within the population. On the other hand, when $R_0$ is below $1$, the disease is less likely to establish a large outbreak. We can say that the basic reproduction number ($R_0$) is an important parameter in epidemiology as it helps to determine the conditions under which an epidemic can take hold and spread throughout a population.

\subsubsection{Critical infection rate ($\beta_{critical}$) }

The critical infection rate ($\beta_{critical}$) is the value of the infection rate ($\beta$) at which the basic reproduction number ($R_0$) becomes greater than $1$. In the context of distance-based epidemic spreading on a Generalized Network Model of Networks (GNMN) using a complex network approach, the critical infection rate can be derived below. From Eq. \eqref{e6}, the basic reproduction number $R_{0}$ can be expressed as.
   \begin{equation*}
    R_0= \frac{\beta}{\mu}\frac{1}{\langle k \rangle}\sum_{i=1}^Ne^{\big (-\frac{d_{T}^2}{2\sigma^2} \big )}\frac{1}{N}\sum_{j=1}^Ne^{\big (-\frac{d_{ij}^2}{2\sigma^2} \big )}
\end{equation*}
As mentioned earlier, critical infection rate $\beta_{critical}$ can be obtained when $R_0 >1$ so it can be expressed as
\begin{equation}
    R_0 >1  \nonumber
\end{equation}
\begin{equation}
 \frac{\beta}{\mu}\frac{1}{\langle k \rangle}\sum_{i=1}^Ne^{\big (-\frac{d_{T}^2}{2\sigma^2} \big )}\frac{1}{N}\sum_{j=1}^Ne^{\big (-\frac{d_{ij}^2}{2\sigma^2} \big )} > 1 \nonumber
\end{equation}

\begin{equation}
\beta > \frac{\mu \langle k \rangle} {\sum_{i=1}^Ne^{(-\frac{d_{T}^2}{2\sigma^2})}\frac{1}{N}\sum_{j=1}^Ne^{(-\frac{d_{ij}^2}{2\sigma^2})}}  \nonumber
\end{equation}

\begin{equation} 
\beta_{critical} > R_{critical} \frac{\mu \langle k \rangle} {\sum_{i=1}^Ne^{(-\frac{d_{T}^2}{2\sigma^2})}\frac{1}{N}\sum_{j=1}^Ne^{(-\frac{d_{ij}^2}{2\sigma^2})}} \label{e7}
\end{equation} 
Equation \eqref{e7} gives the critical infection rate ($\beta_{critical}$) at which the disease has the potential to establish a significant outbreak within the population, taking into account the distance-based effects on the spread of infection within a GNMN.

Epidemic size can be calculated as,\\
\begin{equation} \label{size_E}
    Epidemic\enspace Size = \sum_{i=1}^N \int_{0}^{T} I_i(t)dt \\
\end{equation}
Epidemic size can be explained as the occurrence of the total infection during the epidemic process.

While spreading speed of infection in a geographical region will be,
\begin{equation} \label{Speed_E}
    Spreading\enspace Speed= \frac{1}{N}\sum_{i=1}^N\frac{dI_i}{dt}\frac{1}{A} \\
\end{equation}
Epidemic Spreading speed can be defined as the average infection rate in a particular geographical area.

\section{Results and Analysis}
%\subsection{Materials and methods}
\vspace{-0.4cm}
This section presents the collection and analysis of real-world data on the human population across various Indian states, along with migration patterns to illustrate human connectivity using GNMN. Furthermore, COVID-19 data are incorporated to study the spread of the disease through human movement within the GNMN framework. 

\subsection{COVID-19 Spreading in various Indian States}

\begin{figure}[!htb]
\begin{tabular}{cc}
\subfloat[Uttar Pradesh]{\includegraphics[width= 0.45\columnwidth]{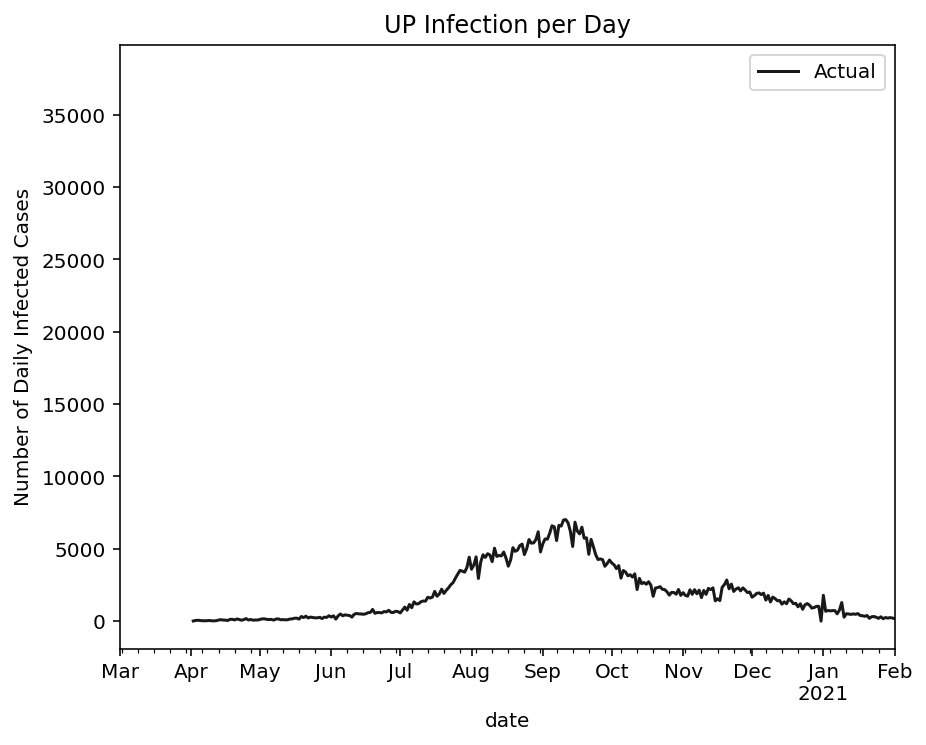}} &
%\subfloat[Delhi]{\includegraphics[width= 0.33\columnwidth]{INDIA_COVID_DATA/DL.png}} &
\subfloat[Maharashtra]{\includegraphics[width= 0.45\columnwidth]{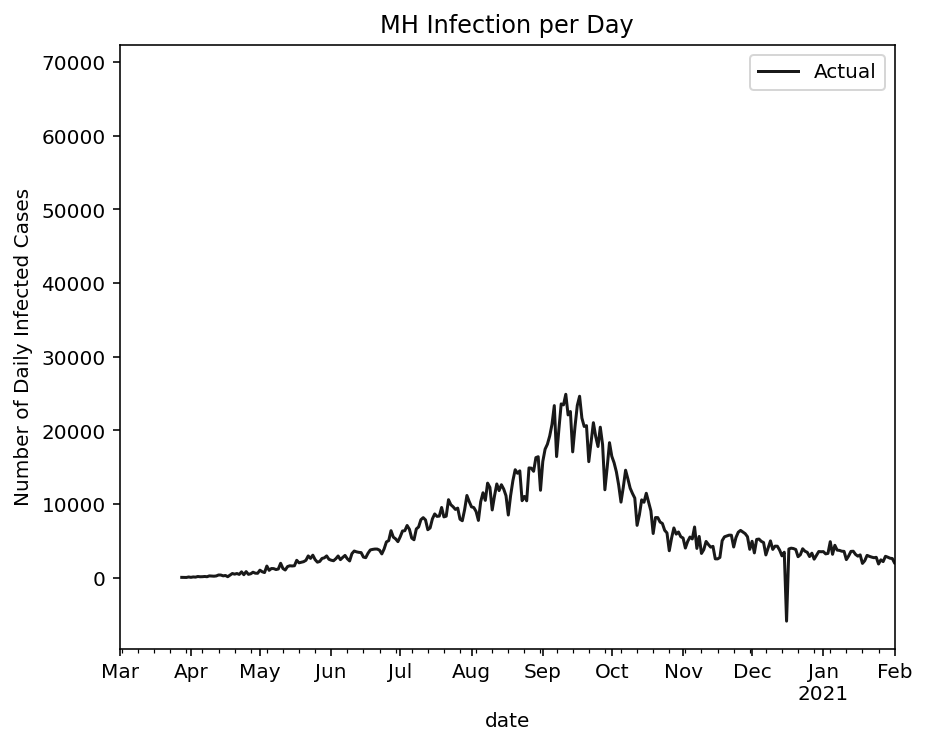}} \\
\end{tabular}
\caption{{Number of Daily infection in Indian States from March 1, 2020 to February 1, 2021}}
\label{State_infection}
\end{figure}

COVID-19 data is collected from March 1, 2020, to February 1, 2021, capturing the spread of infection primarily driven by the movement of migrant workers between states, particularly Maharashtra and Uttar Pradesh. These states experienced a significant impact due to human mobility. Figure \ref{State_infection} illustrates the daily infection trends in Uttar Pradesh and Maharashtra, which had the highest number of migrant workers according to the 2011 Census of India.

The spread of COVID-19 infections in India began in March 2020 but escalated significantly after April 2020. The infection rate remained at its peak between July and December 2020. In Uttar Pradesh, the infection peaked during August, September, and October, after which it started to decline, as illustrated in Figure \ref{State_infection}(a). Maharashtra experienced the most severe outbreak, reaching a peak of 275,000 daily infections between September and October 2020. After December 2020, the daily infection rate stabilized, as shown in Figure \ref{State_infection}(b).

\subsection{Analyzing the effect of network parameters on the degree distribution of the sampled data}

With the help of simulation, we aim to replicate the actual spread of COVID-19 based on real-world data. The latest COVID-19 data are sourced from the official Indian government website \cite{coronaoutbreak}, while the Indian population statistics are obtained from \cite{Indian_Population}. Migration data from various states are collected from \cite{Indian-migration}, and information on human travel speed and resting time is derived from \textbf{Brightkite} \cite{cho2011friendship}. COVID-19 data, including \textbf{Confirmed Infections} and \textbf{Recovered Cases}, is gathered and updated daily for each Indian state from March 1, 2020, to February 1, 2021. Since migration data are based on the 2011 Census, the Indian population data is also collected state-wise from the same census. To account for the increasing population and migration trends over time, a sampling approach is applied. The sampling is conducted after collecting data from both Indian states and migration records, given the large and outdated population data from 2011. An error margin of 1\% with a confidence level of 99\% is used, considering a $p$-value of 3\%. The Biostatistics model proposed by Daniel \cite{daniel2018biostatistics} is applied to sample the human population accordingly and is provided in Eq. \eqref{sampling}.
\begin{eqnarray}
n = \frac{[z^2 * p * (1 - p) / e^2]}{ [1 + (z^2 * p * (1 - p) / (e^2 * N))]} \label{sampling}
\end{eqnarray} 
Where $z$ = 2.576 for a confidence level of 99\%, $p$ = proportion of the population involved in the sampling, $N$ = population size, $e$ = margin of error.

In Table \ref{Popu_table_data}, the total number of actual and migrated populations for both the original and sampled population are presented. The analysis focuses on two states; Maharashtra and Uttar Pradesh, as they were among the most affected during the COVID-19 pandemic. Simulations are conducted to generate networks based on the sampled human population and migration data. Using the data from Tables \ref{Popu_table_data} and \ref{Brightkite_table_data}, networks are constructed following the GNMN model, which examines the time of rest $t_{rest}$, and the connectivity region $r$, within a defined spatial area.

\begin{table}[h!] 
\caption{Actual and sampled State Population of India according to census of 2011 \cite{Indian_Population, Indian-migration}.}
\begin{center}
\begin{tabular}{|p{2cm}|p{2.0cm}|p{2.0cm} |p{2.0cm}|p{2.0cm}|p{2.0cm}|}
\hline
\multirow{2}{*}{\textbf{Indian States }} & \multicolumn{2}{c|}{\textbf{Total Population}} & \multicolumn{2}{c|}{\textbf{Migrated Population}} \\ 
 & Original & Sampled & Original & Sampled \\ \hline

Maharashtra  & 112374333 & 7724  & 57376776 & 3944   \\ \hline
Uttar Pradesh & 199812341 & 7724  & 56452083 & 2182  \\ \hline

\end{tabular}
\end{center}
\label{Popu_table_data}
\end{table}

\begin{table}[h!]  
\caption{Human movement speed and their time of rest $t_{rest}$ according to data of \textbf{Brightkite} Dataset \cite{arquam2020epidemic}}
\begin{center}
\begin{tabular}{|p{4.0cm}||p{4.0cm}||p{3.0cm} ||p{2.0cm}|}
\hline
\textbf{Name of Parameters } & \textbf{Value}  \\ \hline
$t_{rest}$  &  2 days to 115 days  \\ \hline
Connectivity radius($r$) & 2m \\ \hline
Velocity($V$)   & 1m/s to 50m/s  \\ \hline
Area & 25000 $\times$ 25000 $m^2$  \\ \hline
Length of side ($a$) &  25000m  \\ \hline
\end{tabular}
\end{center}
\label{Brightkite_table_data}
\end{table}

\begin{figure}[!htb]
\begin{tabular}{ccc}
\subfloat[Migration from Delhi]{\includegraphics[width= 0.33\columnwidth]{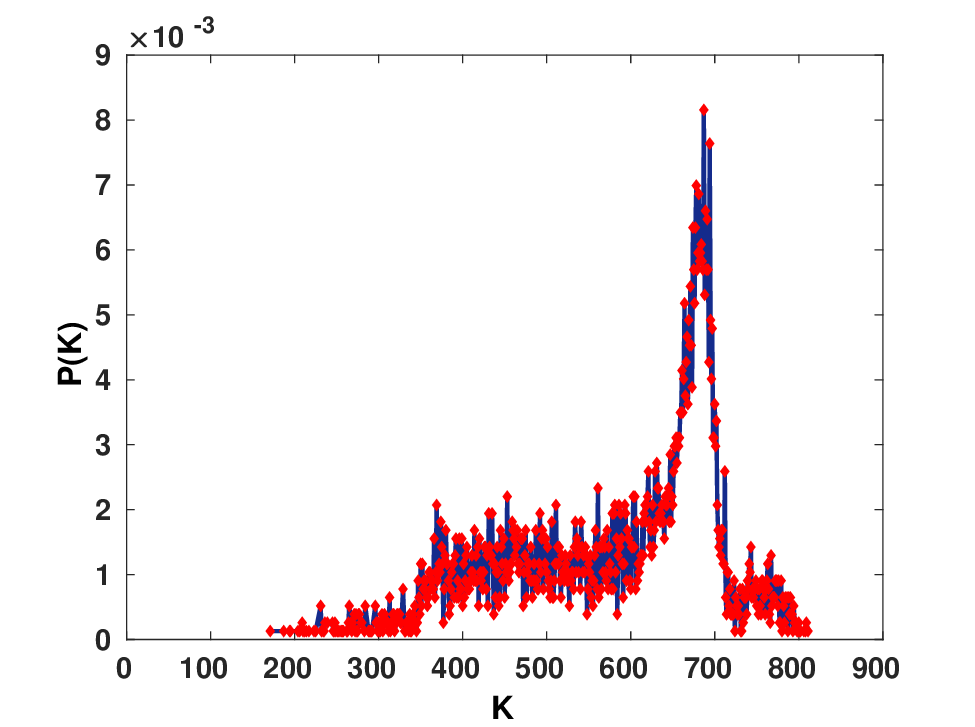}} &
\subfloat[Migration from Maharashtra]{\includegraphics[width= 0.33\columnwidth]{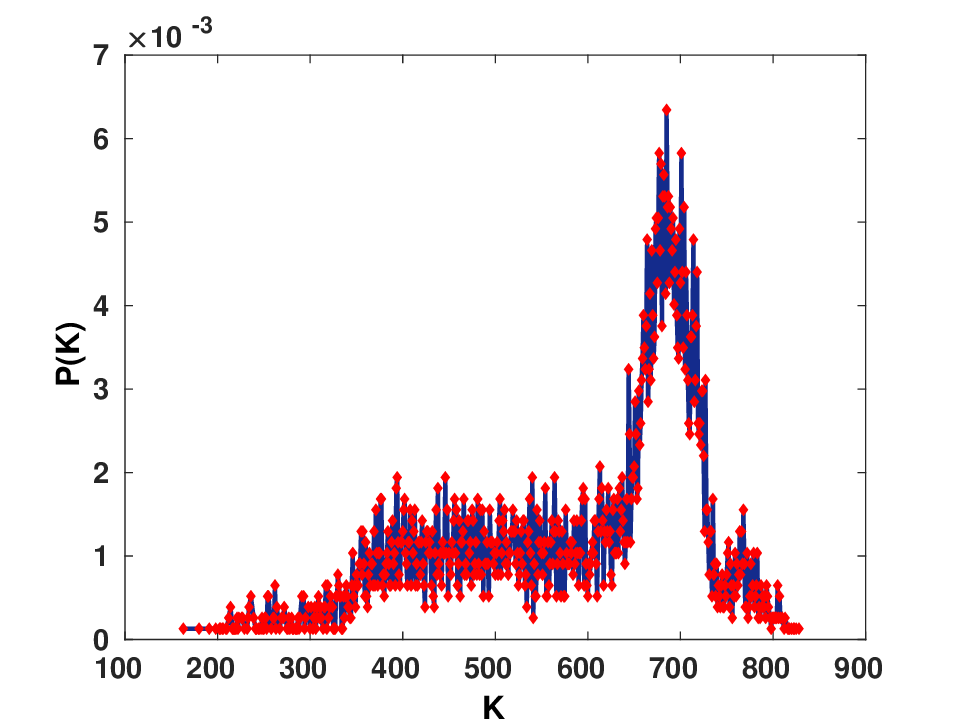} } &
\subfloat[Migration from Gujrat]{\includegraphics[width= 0.33\columnwidth]{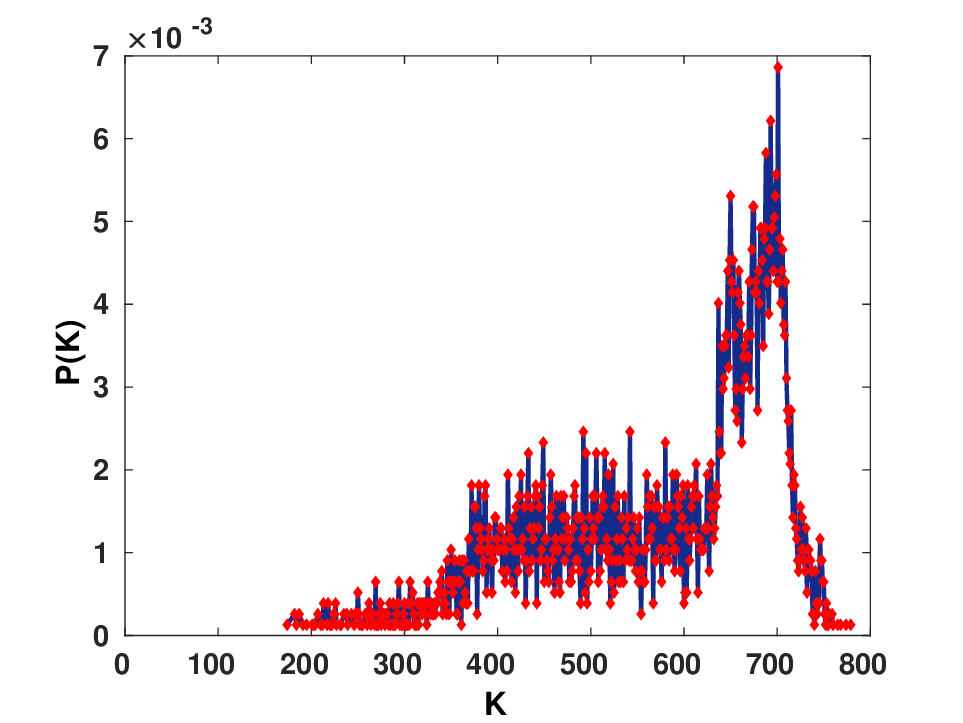}} \\
\end{tabular}
\caption{{Degree Distribution of Human Connection Network of Uttar Pradesh considering movement of migrated population from various states}}
\label{UP}
\end{figure}

The static population and migrating population from various states moving to Uttar Pradesh are utilized to construct networks. The GNMN model is employed to establish a network of states with the highest migration influx into Uttar Pradesh. Figure \ref{UP} illustrates the degree distribution of human contact networks in Uttar Pradesh, incorporating the number of migratory individuals, their rest duration ($t_{rest}$), and the number of halting points. The higher-degree nodes corresponding to Delhi, Maharashtra, and Gujarat fall within the ranges of $600–700$, $650–750$, and $625–725$, respectively, as depicted in Figure \ref{UP}. The probabilities of high-degree nodes for these states are $0.0082$ for Delhi, $0.0063$ for Maharashtra, and $0.007$ for Gujarat.

\begin{figure}[!htb]
\begin{tabular}{ccc}
\subfloat[Migration from Andhra Pradesh]{\includegraphics[width= 0.33\columnwidth]{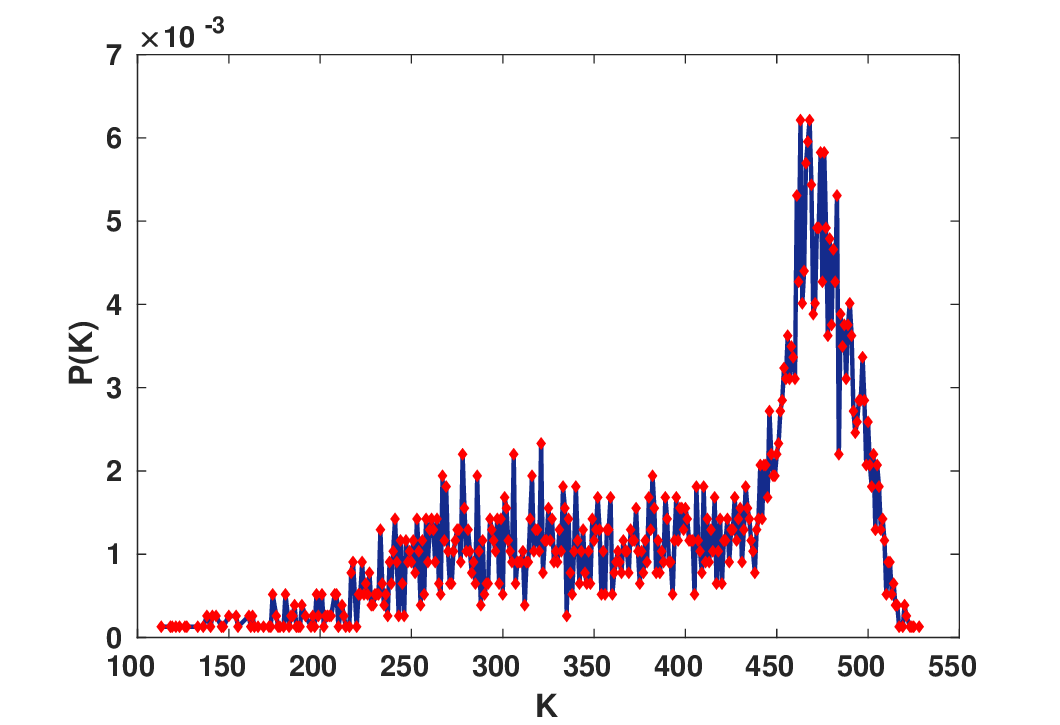}} &
\subfloat[Migration from Bihar]{\includegraphics[width= 0.33\columnwidth]{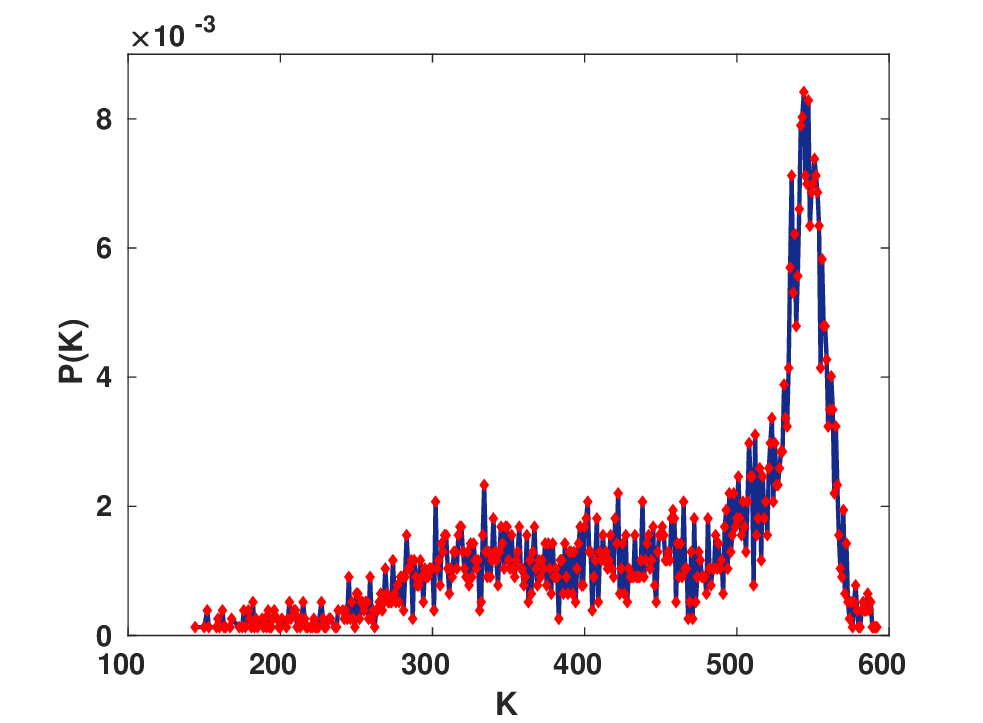}} &
\subfloat[Migration from Uttar Pradesh]{\includegraphics[width= 0.33\columnwidth]{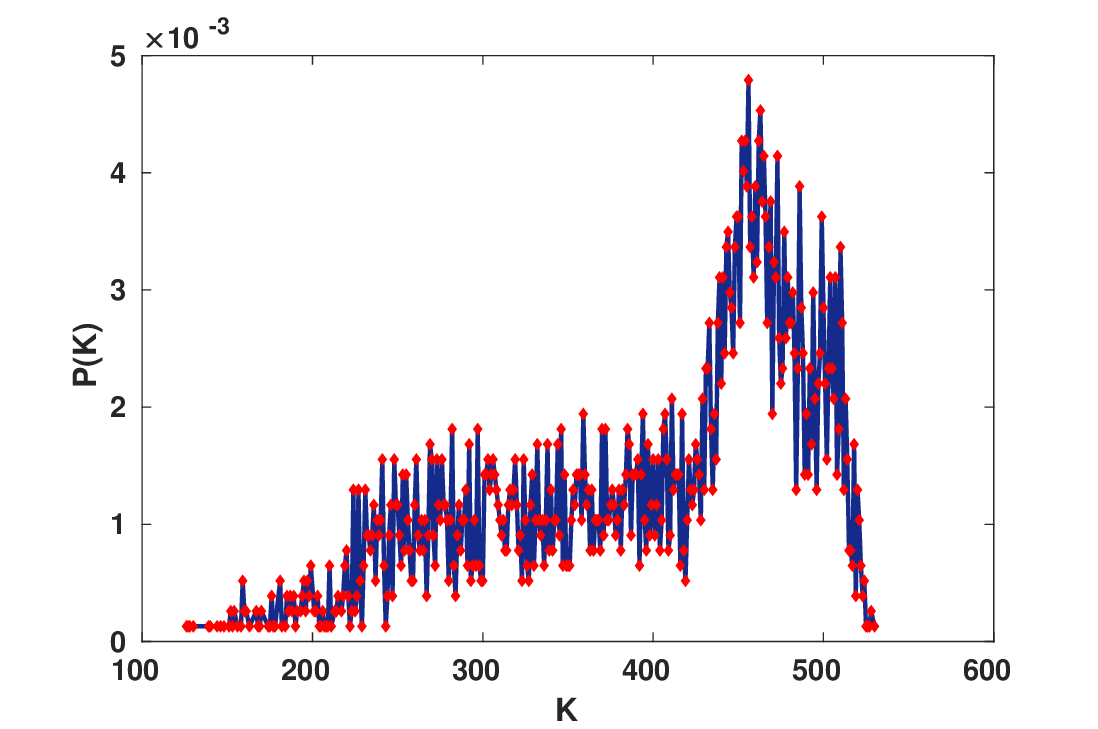}} \\
\end{tabular}
\caption{{Degree Distribution of Human Connection Network of Maharashtra considering movement of migrated population from various states}}
\label{MH}
\end{figure}

Networks are constructed using static and migrated populations from various states moving to Maharashtra. The states with the highest migration influx into Maharashtra are selected for network formation using the GNMN model. The degree distribution of these networks is depicted in Figure \ref{MH}, where variations in the distribution are influenced by the number of migrating individuals, their rest time ($t_{rest}$), and the number of stopping points. From Figure \ref{MH}, it is observed that higher-degree nodes fall within the ranges of $450–500$ for Andhra Pradesh, $500–600$ for Bihar, and $400–520$ for Uttar Pradesh. The probabilities of high-degree nodes for these states are $0.006$ for Andhra Pradesh, $0.008$ for Bihar, and $0.0049$ for Uttar Pradesh.

From the aforementioned plots for degree distribution, it is evident that each network's degree distribution differs due to node mobility, which varies with velocity ($V$) and rest time ($t_{rest}$). Consequently, each graph exhibits distinct structural properties, leading to variations in the average degree. We hypothesize that when nodes move with random velocities and shorter $t_{rest}$ times within a given spatial region, the network becomes sparser, resulting in fluctuating degree distributions. The density of the node is determined by the duration of $t_{rest}$. A longer $t_{rest}$ means that other nodes from different locations are more likely to be drawn into the connectivity region ($r$) of the stationary node, thus influencing the formation of the network.

\subsection{Visualization of infection spread using GNMN model}

\begin{figure}[!htb]
\begin{tabular}{c}
\includegraphics[width= 0.95\columnwidth]{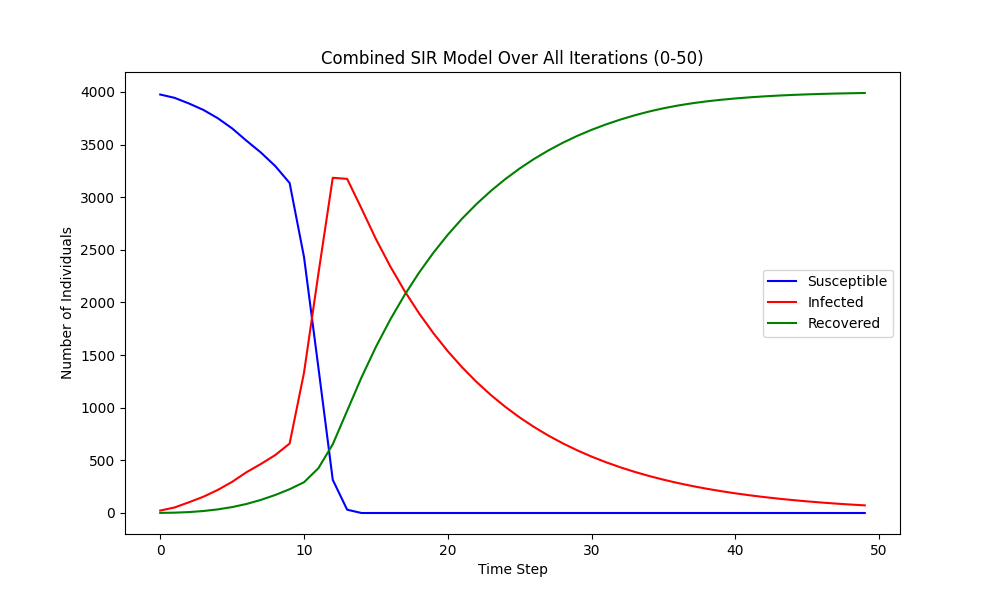}
%\subfloat[Maharashtra]{\includegraphics[width= 0.45\columnwidth]{Maharashtra/R0_vs_time.png}} \\
\end{tabular}
\caption{{The dynamics of Epidemic spreading is plotted to show the change in population in different compartments of the total population}}
\label{SIR_Combined}
\end{figure}

The infection propagation is modeled and visualized in Figure \ref{SIR_Combined} to analyse the dynamics of infection spread within the human contact network. The SIR model is used to track population changes across compartments over time, with a synthetic network consisting of $4000$ individuals. When the epidemic begins in the network, assuming GNMN as the underlying topology, the infection spreads slowly for the first nine time steps (as shown in Fig. \ref{SIR_Combined}). After this initial phase, the infection rate rises sharply, reaching 3,160 infected individuals by the $12^{th}$ time step. Concurrently, the susceptible population declines gradually at first but drops steeply after nine time steps. Following this peak, the infection rate starts to decline, and from the $14^{th}$ time step onward, it follows a monotonous decreasing trend. The recovery process accelerates after the $10^{th}$ time step, and by the $50^{th}$ time step, the infection is completely eradicated, rendering the entire population infection-free.

\begin{figure}[!htb]
\begin{tabular}{cc}
\subfloat[Uttar Pradesh]{\includegraphics[width= 0.45\columnwidth]{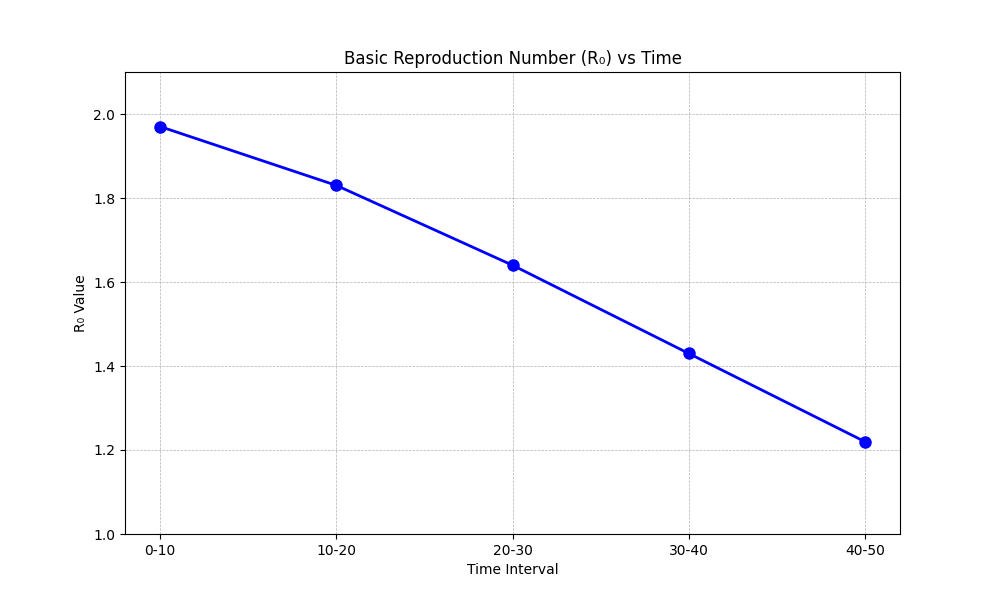}} &
\subfloat[Maharashtra]{\includegraphics[width= 0.45\columnwidth]{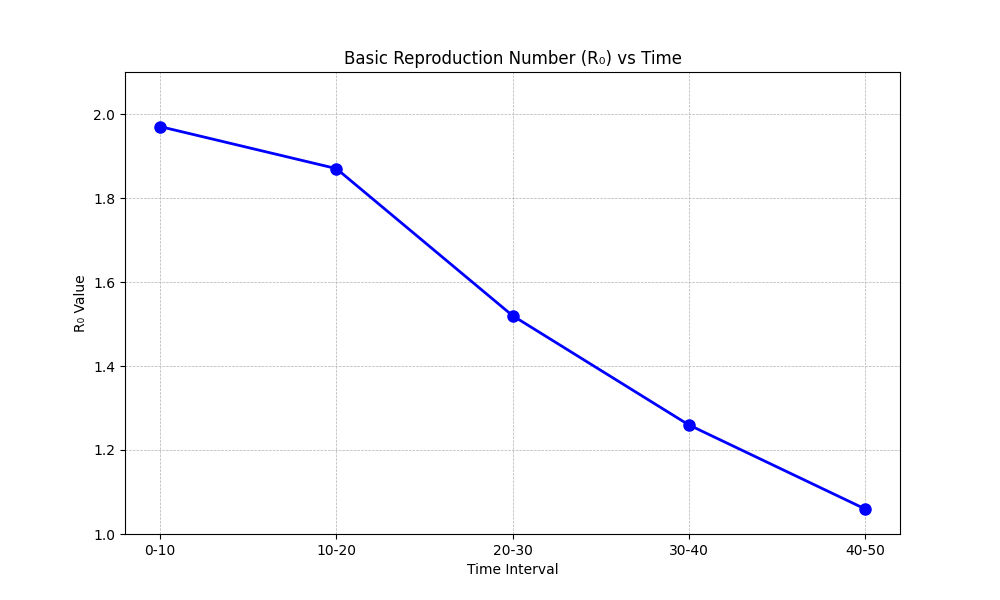}} \\
\end{tabular}
\caption{{$R_0$ is plotted to show the basic reproduction number of COVID-19 in the state of Uttar Pradesh and Maharashtra}}
\label{R_0}
\end{figure}

The basic reproduction rate $R_0$ for real data from Uttar Pradesh and Maharashtra is plotted alongside the corresponding $(R_0)$ values from the proposed model in Figure \ref{R_0}. To generate the $R_0$ plot for the proposed model, parameters from Table \ref{Brightkite_table_data} were used. The plot exhibits a similar curve pattern to the real data. In the proposed model, $R_t$ begins at $3.7$ and declines to $0.4$ after $100$ time steps. In India's real data,$R_0$ started at $3$ in March 2020 and gradually decreased to approximately $1$ by February 2021.

\begin{figure}[!htb]
\begin{tabular}{cc}
\subfloat[Uttar Pradesh]{\includegraphics[height = 4 cm, width= 0.48\columnwidth]{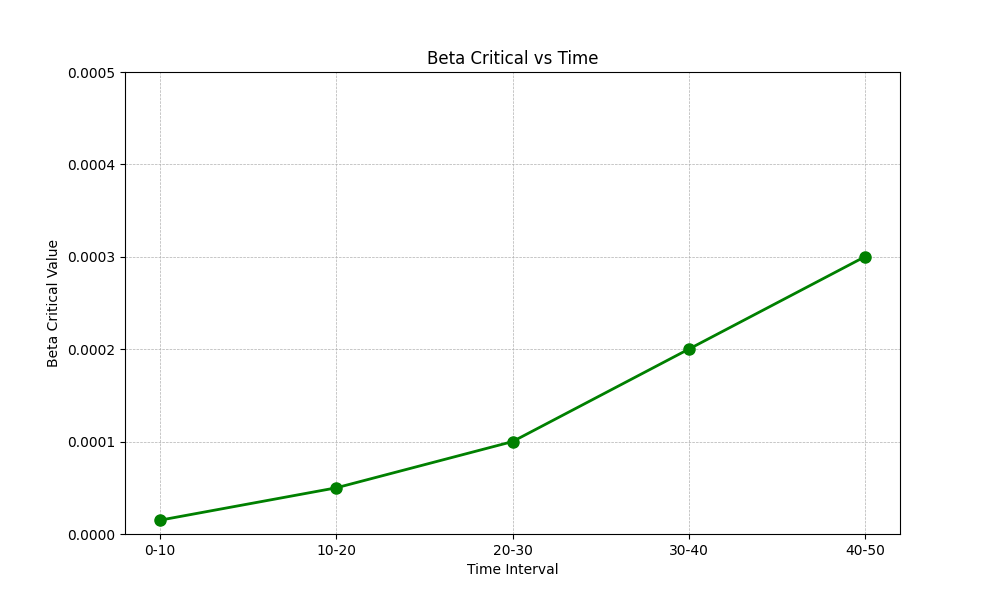}} &
\subfloat[Maharashtra]{\includegraphics[height = 4 cm, width= 0.48\columnwidth]{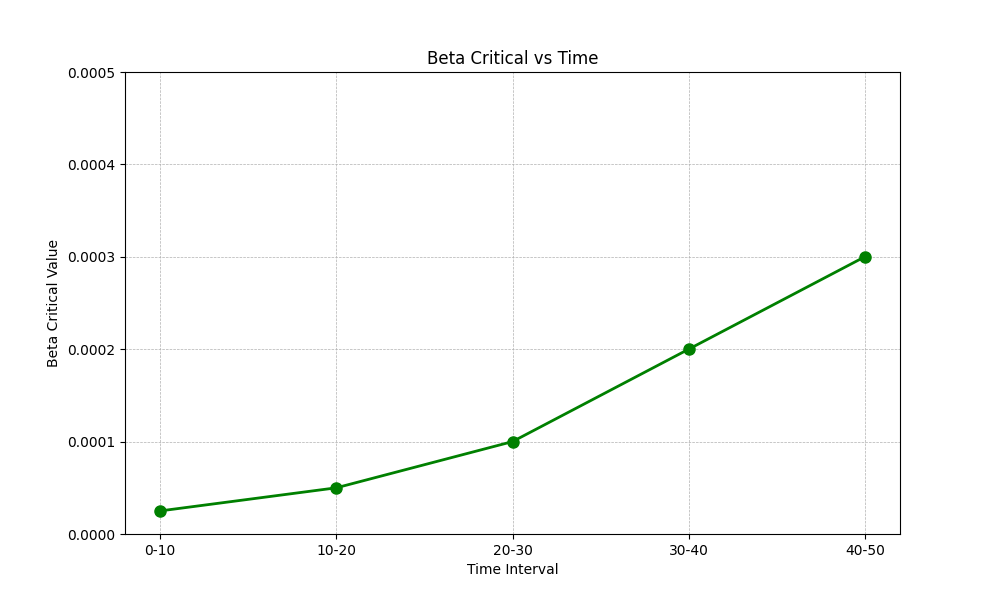}} \\
\end{tabular}
\caption{{$\beta_{Critical}$ is plotted to show the time of outbreak of COVID-19 in the state of Uttar Pradesh and Maharashtra}}
\label{Beta_Crit}
\end{figure}

The effect of the spread rate over time is illustrated in Figure \ref{Beta_Crit}. At time step $10$, the spreading rate reached $0.000028$ for both Uttar Pradesh and Maharashtra. Beyond this point, the epidemic spreading rate continued to rise. This critical threshold marks the onset of the outbreak. The simulation was conducted for $40$ time steps, with the maximum attainable spreading rate reaching 
$0.0003$ for both states.

\begin{figure}[!htb]
\begin{tabular}{cc}
\subfloat[Uttar Pradesh]{\includegraphics[height = 4 cm, width= 0.5\columnwidth]{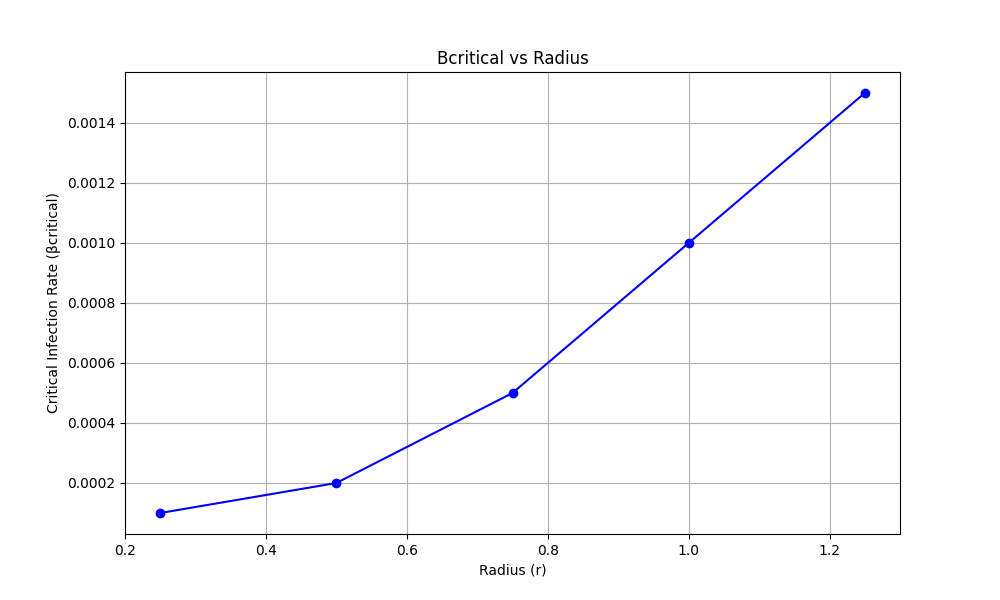}} &
\subfloat[Maharashtra]{\includegraphics[height = 4 cm, width= 0.5\columnwidth]{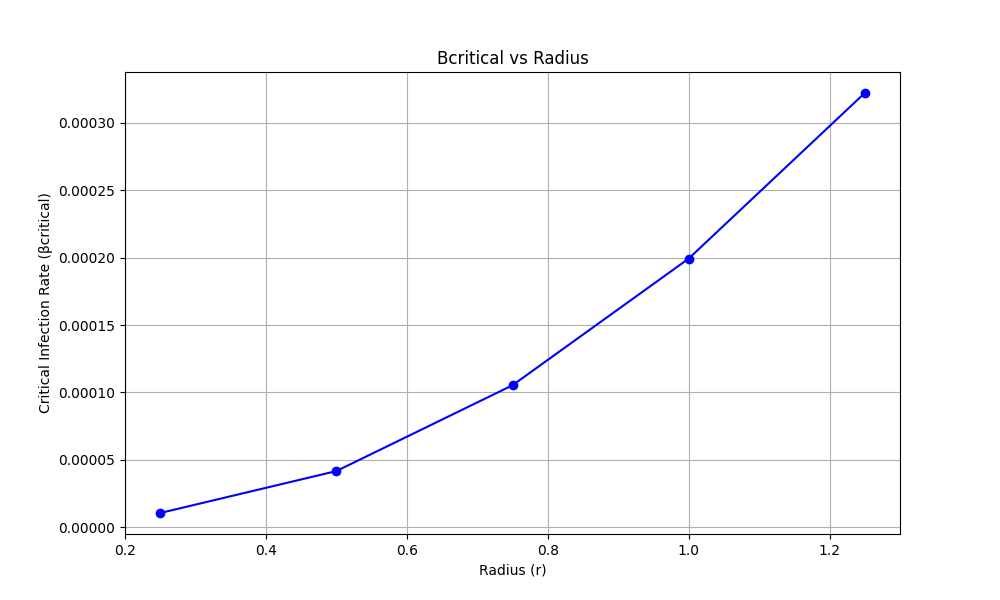}} \\
\end{tabular}
\caption{{$\beta_{critical}$  is plotted against varying radius of connectivity to show the outbreak of COVID-19 on the network using GNMN}}
\label{beta_c_sampled data}
\end{figure}

Figure \ref{beta_c_sampled data} illustrates the relationship between the spreading rate and the varying radius of connectivity. The result indicates that the spreading rate is highly impacted by varying radius of connectivity. At time step $10$, the spreading rate reaches $0.000028$ for both Uttar Pradesh and Maharashtra, marking a critical threshold that triggers the epidemic outbreak. Beyond this point, the spreading rate continues to rise. The simulation runs for $40$ time steps, with the maximum achievable spreading rate reaching $0.0003$ for both states.

\section{Conclusions and Future Scopes}

In this manuscript, we analyzed the importance of human contact and distance in the spread of disease, specifically in relation to the COVID-19 pandemic. By incorporating distance parameterization into the classical SIR model, the critical threshold of epidemic spreading is derived, such as the basic reproduction number $R_0$ and the critical infection rate $\beta_{critical}$, and how they are influenced by human movement. The model also accurately portrays the spread of infection in a human connection network. The findings challenge the traditional depiction of infection spread and prove that the topology of the connection strongly influences human pandemic mobility. The model proposed in this paper helps to understand how the connecting distance constraining factors can help mitigate the spread of disease and control an outbreak. This research examines how people movement and spatial distance affect an area’s epidemiological activity. The model is based on a distance-based version of SIR network model, constructed on the notion of distance-based human contact network. The evidence is consistent with the theory that mobility and spatially structured interaction is essential for the spread of the COVID-19 virus. It also supports the idea that during pandemics, how people move around the world strongly affects infection spread.

In the future work, behavioral patterns, vaccination effects, and environmental factors can be incorporated to enhance the model further. The integration of multi-layered networks can improve forecasting for multi-faceted epidemics because of different human interactivity aspects. Mobility trends can be utilized to design more effective intervention strategies through AI and machine learning. Broadening the scope of the study to include various infectious diseases and geographical regions will strengthen its applicability and accuracy in disease prevention and control.

\bibliographystyle{unsrt} 
\bibliography{Epid-distan-mobile}

%\vspace{-.8cm}
%\bibliographystyle{unsrt}
%\bibliography{Blockchain}
%\vspace{-.8cm}
\end{document}